\documentclass[copyright,creativecommons]{eptcs}
\usepackage[latin1]{inputenc}

\usepackage{upgreek,latexsym, amssymb, amsmath, amsfonts, mathrsfs, bussproofs}

\usepackage{xcolor}

\newtheorem{theorem}{Theorem}
\newtheorem{lemma}{Lemma}
\newtheorem{definition}{Definition}
\newtheorem{example}{Example}

\usepackage{latexsym}
\usepackage{amsmath}
\usepackage{amssymb}

\newcommand{\com}{\newcommand}
\com{\pgt}[1]{{\tt #1}}

\newcommand{\lsem}{\mbox{$\lbrack\hspace{-0.3ex}\lbrack$}}
\newcommand{\rsem}{\mbox{$\rbrack\hspace{-0.3ex}\rbrack$}}

\newcommand{\nats} {{\mathbb N}}

\newcommand{\fl}{\noindent}
\newcommand{\hair}{\hspace{3mm}}
\newcommand{\vair}{\vspace{3mm}}

\newcommand{\be}{\begin{enumerate}}
\newcommand{\ee}{\end{enumerate}}

\newcommand{\bi}{\begin{itemize}}
\newcommand{\ei}{\end{itemize}}

\newcommand{\ba}{\begin{array}}
\newcommand{\ea}{\end{array}}

\newcommand{\bdfn}{\begin{definition}}
\newcommand{\edfn}{\end{definition}}

\newcommand{\bex}{\begin{example}}
\newcommand{\eex}{\end{example}}

\newcommand{\bt}{\begin{tabular}}
\newcommand{\et}{\end{tabular}}

\newcommand{\sempar}[1]{\mbox{\lsem{\tt #1}\rsem}}

\newcommand{\ldec}{\mbox{$\mathbf{\{\hspace{-1mm}\{}$}}
\newcommand{\rdec}{\mbox{$\mathbf{\}\hspace{-1mm}\}}$}} 

\newcommand{\decidepar}[1]{\mbox{\ldec\rm{#1}\rdec}}

\newcommand{\bind}[2]{#1\mapsto#2}

\newcommand{\bp}{\begin{program}}
\newcommand{\ep}{\end{program}}

\newcommand{\programenvironment}{\programmode%
	\def\par{\leavevmode\endgraf}\obeylines\nobreak%
	\programmode}

\newcommand{\programmode}{\normalsize\tt
	\catcode`\_=12 \catcode`\?=12 \catcode`\.=12 \catcode`\,=12
	\catcode`\;=12 \catcode`\:=12 \catcode`\@=12 \catcode`\~=12
        \catcode`\#=12 \catcode`\&=12      
	\obeyspaces\frenchspacing}%

\newenvironment{programintext}{\programenvironment}{}

\newenvironment{program}{\setlength{\partopsep}{0mm}\setlength{\topsep}{0mm}
	\begin{trivlist}\item[]
\begin{minipage}{0.90\textwidth}
	\vspace{1mm}
	\begin{programintext}
	}{\end{programintext}
	\vspace{1mm}
	\end{minipage}
	\end{trivlist}
	\noindent}

{\catcode`\^^I=\active \obeyspaces\global\def^^I{ }}
{\obeyspaces\global\let =\ } 
{\catcode`\`=\active \gdef`{\relax\lq}}

\com{\pf}[2]{#1 \rightarrow #2_{\bot}}

\newcommand{\prog}{{\tt p}}
\newcommand{\strue}{{\tt True}}
\newcommand{\sfalse}{{\tt False}}

\newcommand{\tpx}{{\cal T}^{{\tt p},x}}

\title{Cons-free Programs and Complexity Classes between 
       {\sc logspace} and {\sc ptime}}
\author{
\phantom{ABCDEFGHIJKLMNOP}
\and
Neil D. Jones
\institute{Computer Science Department\\\
University of Copenhagen}
\email{neil@diku.dk}
\and
\phantom{ABCDEFGHIJKLMNOP}
\and
Siddharth Bhaskar
\institute{Computer Science Department\\\
University of Copenhagen}
\email{sbhaskar@di.ku.dk}\and
Cynthia Kop
\institute{Department of Software Science\\
Radboud University. Nijmegen}
\email{C.Kop@cs.ru.nl}
\and
Jakob Grue Simonsen
\institute{Computer Science Department\\\
University of Copenhagen}
\email{simonsen@di.ku.dk}
}
\def\titlerunning{Cons-free Programs and Complexity Classes between        {\sc logspace} and {\sc ptime}}
\def\authorrunning{Jones, Bhaskar, Kop, Simonsen
}

\begin{document}

\setcounter{tocdepth}{5}
\setcounter{theorem}{0}
\setcounter{lemma}{0}
\setcounter{definition}{0}



\maketitle

\begin{abstract}
Programming language concepts are used to give some new perspectives on  a long-standing open problem: is {\sc logspace} = {\sc ptime} ?
\end{abstract}

\section*{Introduction}

 ``P =? NP''    is an
archetypical  question in computational complexity theory, unanswered
 since its formulation in the 1970s. The question: Is the computional power of polynomially time-bounded programs increased  by adding
the ability to ``guess'' (i.e., nondeterminism) ?
This is interesting because ``polynomial time'' is a plausible candidate for ``feasibly  solvable''.

Perhaps the second most important question is ``L =? P'': whether $\mbox{\sc logspace}  =  \mbox{\sc ptime}$. Here L is the set of problems solvable by
{\em cursor programs}. These also run in polynomial time, but have no rewritable 
storage\footnote{One take: a  cursor program is a multihead two-way read-only finite automaton. A more classical but equivalent version:  a 2-tape Turing machine with $n$-bit read-only input tape 1, that uses at most  
$O(\log n)$ bits of storage space on read-write tape 2.}.
Both  questions
remain open since Cook and Savitch's pathbreaking papers  in the 1970s
\cite{Cook:71:CharacterizationOfPushdown,DBLP:journals/jcss/Savitch70}.

We investigate   the question ``L =? P'' from the viewpoint of {\em functional programming languages}: a different viewpoint than Turing machines. The link is 
earlier characterisations of L and P by ``cons-free'' programs  \cite{Jones:97:ComputabilityComplexity,logspace,jfp}. 
The net result:  a deeper and finer-grained analysis,  illuminated by perspectives both from
  programming languages and complexity theory.
 
Some new definitions and theorems  give fresh perspectives on the  question  L =? P. We use programs to define and study complexity classes  between the two. 
By \cite{Jones:97:ComputabilityComplexity,logspace,jfp} cursor programs exactly capture the problem class L; and cursor programs with {\em recursive function definitions} exactly capture the problem class P. A drawback though is that recursive
cursor programs  {\em can run for exponential time}, even though they exactly capture the {\em decision problems} that can be solved  in polynomial time by Turing machines.

{\bf The goal of this paper} is to better understand the problems  in the interval between classes L and P.  
Problem class NL is  already-studied in this interval, and it is the logspace analog of similar  long-standing open problems. Kuroda's two ``LBA problems'' posed in 1964 \cite{Kuroda}: (1) Is {\sc dspace($n$)} =?  {\sc nspace($n$)} and (2) Is {\sc nspace($n$)} closed under complementation? After both stood unresolved for 23 years, (2) was finally 
answered "yes" (independently in 1987)  by Immerman and by Szelepcs{\'{e}}nyi \cite{Immerman,Szelepcsenyi}: NL and larger nondeterministic space classes 
(with constructive bounds) are closed under complementation.\footnote{Kuroda's other LBA problem 
{\sc dspace($n$)} =?  {\sc nspace($n$)} is still open, as well as the question L =? NL.}
 
 We study the problems solvable by an in-between class CFpoly: recursive cursor programs that {\em run in polynomial time}.  Recursion is in some sense orthogonal to the ability to make nondeterministic choices, i.e., to ``guess''. The class CFpoly seems more natural than NL from a programming perspective.

\section{Complexity by Turing machines and by programming languages}

\subsection{Overview
}

Let $X \subseteq \{0,1\}^*$ be a set of bit strings. The {\em decision problem} for $X$: given $x \in \{0,1\}^*$, to decide whether or not $x \in X$. We say that $X$ is in {\sc ptime} iff it is decidable by a 1-tape deterministic Turing machine 
that runs within polynomial  time $O(n^k)$ [here $n = |x|$ is the length of its input $x$, and $k$ is a constant independent of $x$]. 
Further,  $X$  is in {\sc logspace} iff it is decidable by a 2-tape Turing machine that uses at most  
$O(\log n)$ bits of storage space on tape 2, assuming its $n$-bit input is given on read-only tape 1. 
Both problem  classes  are 
of practical interest as they are decidable by programs with running times  bounded by polynomial functions of their input lengths. The essential difference is the amount of allowed storage space.
These classes are invariant across a wide range of variations among computation models, and it is easy to see that $\mbox{\sc logspace}  \subseteq  \mbox{\sc ptime}$.

However, the question: is $\mbox{\sc logspace}  \subsetneq  \mbox{\sc ptime}?$ has stood open for many years.
\vair

\fl{\bf Programs and problem decision.}
Semantics: a program computes a (partial!) function from bit strings to bits:
$$\lsem{\tt p}\rsem:\{0,1\}^* \to  \{0,1\}\cup\{\bot\}$$
Program semantics is call-by-value; and $\lsem{\tt p}\rsem(x) = \bot$ means: {\tt p} does not terminate on input $x$.
\vair

\fl A set $X \subseteq  \{0,1\}^*$ is {\em decided} by a  program {\tt p} if
{\tt p} terminates on all inputs $x \in \{0,1\}^*$, and for any  $x$, 
$$
\lsem{\tt p}\rsem(x) = \left\{ \ba{ll} 1 & \mbox{if $x\in X$} \\
                                                      0 & \mbox{if $x\notin X$} 
                                            \ea
                                   \right.
$$

\fl{\bf Complexity by cons-free programs.}
We use programming languages  (several  in this paper) to explore the interesting boundary zone between the  problem classes $\mbox{\sc logspace}$	 and $\mbox{\sc ptime}$. 
Strong links were established in \cite{logspace,jfp}:
Each class was characterised by a small general recursive functional programming language. The one language (called CF for ``cons-free'') is limited to programs without data constructors.\footnote{``Cons-free'' is not to be confused  with ``context-free''. TR stands for ``tail recursive''.  Data access is {\em read-only} in both of our languages,
so neither  CF nor CFTR is Turing-complete.}  The other,  named CFTR, is identical to  CF but has restricted control, allowing only {\em tail recursive}  calls to defined 
functions.\footnote{These ideas stem from S.\ A.\ Cook's seminal work on complexity classes and pushdown machines \cite{Cook:71:CharacterizationOfPushdown}. Papers
 \cite{logspace,jfp} re-express and adapt Cook's ideas to a cons-free programming context. Paper  \cite{jfp} extends \cite{Cook:71:CharacterizationOfPushdown}, characterising decision powers of  higher-order  cons-free  programs. CFTR is (yet another)  version of the ``cursor programs'' mentioned in the Introduction.} From \cite{jfp} we know:

\begin{theorem}\label{thm-l-equivalence}\hfill $\mbox{\sc logspace} = \{X\subseteq \{0,1\}^* \ |\ \mbox{some CFTR program decides } X  \}$\end{theorem}

\begin{theorem}\label{thm-p-equivalence}\hfill
$\mbox{\sc ptime} = \{X\subseteq \{0,1\}^* \ |\ \mbox{some CF program decides } X  \}$
\end{theorem}

\fl{\bf A compact notation relating programs and problems}

\bdfn
 $\decidepar{L}$ is  the set of all {\em problems} (i.e., sets 
$X \subseteq\{0,1\}^*$) that are decidable by {\rm L}-{\em programs}: 
$$
\decidepar{L} \ \ \stackrel{def}{=}\ \  \{X\subseteq\{0,1\}^*\ | \ \mbox{\ some {\rm L}-program {\tt p} decides\ } X \}
$$
\edfn

\noindent {\bf Theorems \ref{thm-l-equivalence} and   \ref{thm-p-equivalence}} can thus be restated as: \hfill  $\mbox{\sc logspace} = \decidepar{CFTR} \subseteq  \decidepar{CF} = \mbox{\sc ptime}$

\subsection{The cons-free programming language CF
}\label{sec-cf}

All programs are  first-order in this paper,\footnote{The larger language described in  \cite{jfp} encompasses both first and higher-order programs.} and deterministic to begin with (until nondeterminism is expressly added in Section \ref{sec-nondeterminism}).  First, a brief overview (details in  \cite{Jones:97:ComputabilityComplexity,logspace,jfp}):
\bdfn\label{def-cf-syntax}

{\bf Program syntax:} a CF-program is a finite sequence of mutually recursive function definitions. Write CF programs and fragments in {\tt teletype}, e.g., program {\tt p} or expression
{\tt tail(f x y)}. The first definition, 
of form ${\tt f_1\ x = e^{f_1}}$,   
defines the program entry function (with one argument, always named {\tt x}).
A context-free grammar for programs, definitions and expressions {\tt e}:

\bp
p    ::= def | p def     -- Program = sequence of function definitions
def  ::= f x1...xm = e   -- (where m >= 0)                      
e    ::= True | False | [] | xi | base | if e then e else e | f e1...em
base ::= not e | null e | head e | tail e      -- Base function call

\ep

{\bf Data types:} bits and bit lists. Variable and expression values are elements of one of the value sets $\{0,1\}$ or list $\{0,1\}^*$. A string $x \in \{0,1\}^*$  is a bit list, and list $1011$ can be written {\tt [1,0,1,1]} as data value.
{\tt []} is the empty list, and {\tt b:bs} is the result of prepending bit {\tt b} to list {\tt bs}.
We sometimes identify {\tt 0, 1} with {\tt False} and 
{\tt True}, resp.
e.g., in the test position {\tt e0} of an expression
{\tt if e0 then e1 else e2}.

{\bf Expressions:} Expression {\tt e} may be 
a constant; a variable; a base function call 
({\tt not}, {\tt null}, {\tt head}, or {\tt tail});
a conditional {\tt if e0 then e1 else e2};
or a call {\tt f\,e1...er} to some program-defined function {\tt f}. 
A program contains no CONS function or other constructors (hence CF for cons-free).  Thus CF does not have successor, $+,*$ or explicit storage allocators such as {\tt malloc} in $C$, {\tt ::} in ML, or {\tt :} in Haskell.

{\bf Function definition:}  A program-defined definition has form {\tt f\,x1\,x2...xm = e} with $m\geq 0$.  No function may be defined more than once,
left-side variables must be distinct, and any variable appearing on the right side (in {\tt e}) must be one of {\tt x1},\ldots,{\tt xm}.

{\bf Semantics:} The semantics of CF is given by the {\em inference rules} 
in Figure~\ref{fig:inferencerules}. Given a program {\tt p} and input $x$, these rules define a (unique) computation tree that we call $\tpx$. The inference rules define statements 
$\lsem{\tt p}\rsem(x)=v$ (at the root of the computation tree) or ${\tt p},\rho\vdash {\tt e} \to v$.  
The {\tt p} in
${\tt p},\rho\vdash {\tt e} \to v$ is a CF program. The 
{\tt e} is a subexpression of the right side of some function definition. Further,    
$\rho$ is an {\em environment} that binds the current program  variables to respective 
 {\em  argument values} $v_1,\ldots, v_m$; and
 $v$ is the {\em result value} of {\tt e}. Base or defined function calls are evaluated using  call-by-value.
\edfn

\begin{figure}[t]
    \centering
\hair\textbf{Axioms:}

\begin{minipage}{0.22\textwidth}
\begin{prooftree}
\AxiomC{}
\UnaryInfC{$\prog,\rho \vdash {\tt x} \to \rho({\tt x})$}
\end{prooftree}
\end{minipage}
\begin{minipage}{0.22\textwidth}
\begin{prooftree}
\AxiomC{}
\UnaryInfC{$\prog,\rho \vdash {\tt True} \to \strue$}
\end{prooftree}
\end{minipage}
\begin{minipage}{0.22\textwidth}
\begin{prooftree}
\AxiomC{}
\UnaryInfC{$\prog,\rho \vdash {\tt False} \to \sfalse$}
\end{prooftree}
\end{minipage}
\begin{minipage}{0.22\textwidth}
\begin{prooftree}
\AxiomC{}
\UnaryInfC{$\prog,\rho \vdash {\tt []} \to []$}
\end{prooftree}
\end{minipage}
\vspace{10pt}

\textbf{Base functions:}

\begin{minipage}{0.3\textwidth}
\begin{prooftree}
\AxiomC{$\prog,\rho \vdash {\tt e} \to \sfalse$}
\UnaryInfC{$\prog,\rho \vdash {\tt not}\ {\tt e} \to \strue$}
\end{prooftree}

\begin{prooftree}
\AxiomC{$\prog,\rho \vdash  {\tt e} \to \strue$}
\UnaryInfC{$\prog,\rho \vdash {\tt not}\  {\tt e} \to \sfalse$}
\end{prooftree}
\end{minipage}
\begin{minipage}{0.3\textwidth}
\begin{prooftree}
\AxiomC{$\prog,\rho \vdash {\tt e} \to {\tt []}$}
\UnaryInfC{$\prog,\rho \vdash {\tt null}\ {\tt e} \to \strue$}
\end{prooftree}

\begin{prooftree}
\AxiomC{$\prog,\rho \vdash {\tt e} \to v_1{\tt :}v_2$}
\UnaryInfC{$\prog,\rho \vdash {\tt null}\ {\tt e} \to \sfalse$}
\end{prooftree}
\end{minipage}
\begin{minipage}{0.3\textwidth}
\begin{prooftree}
\AxiomC{$\prog,\rho \vdash {\tt e} \to v_1{\tt :}v_2$}
\UnaryInfC{$\prog,\rho \vdash {\tt head}\ {\tt e} \to v_1$}
\end{prooftree}

\begin{prooftree}
\AxiomC{$\prog,\rho \vdash {\tt e} \to v_1{\tt :}v_2$}
\UnaryInfC{$\prog,\rho \vdash {\tt tail}\  {\tt e} \to v_2$}
\end{prooftree}
\end{minipage}
\vspace{10pt}

\textbf{Condition:}

\begin{minipage}{0.45\textwidth}
\begin{prooftree}
\AxiomC{${\tt p},\rho  \vdash {\tt e_0} \to \strue$}
\AxiomC{${\tt p},\rho \vdash {\tt e_1} \to v$}
\BinaryInfC{${\tt p},\rho \vdash {\tt if\ e_0\ then\ e_1\ else\ e_2} \to v$}
\end{prooftree}
\end{minipage}
\begin{minipage}{0.45\textwidth}
\begin{prooftree}
\AxiomC{${\tt p},\rho \vdash  {\tt e_0} \to \sfalse$}
\AxiomC{${\tt p},\rho \vdash {\tt e_2} \to v$}
\BinaryInfC{${\tt p},\rho \vdash {\tt if\ e_0\ then\ e_1\ else\ e_2} \to v$}
\end{prooftree}
\end{minipage}
\vspace{10pt}

\textbf{Function call:}
\vspace{-4mm}
\begin{prooftree}
\AxiomC{${\tt p},\rho \vdash {\tt e}_1 \to w_1$ \quad \dots
        \quad ${\tt p},\rho \vdash {\tt e}_m \to w_m$}
\AxiomC{${\tt p},[\bind{{\tt x1}}{w_1},\dots,\, \bind{{\tt xm}\ }{w_m}] \vdash {\tt e}^{\tt f} \to v$}
\RightLabel{if ${\tt f\,x1 \dots xm = e^f} \in {\tt p}$}
\BinaryInfC{${\tt p},\rho \vdash {\tt f e_1 \dots e_m} \to v$}
\end{prooftree}
\vspace{4pt}

\textbf{Program running:}
\begin{prooftree}
\AxiomC{${\tt p},[\bind{\tt x}{\mathit{input}}] \vdash {\tt e}^{{\tt f}_1} \to v$}
\RightLabel{If ${\tt f}_1\ x = {\tt e}^{{\tt f}_1}$ is the entry function definition of ${\tt p}$}
\UnaryInfC{$\lsem{\tt p}\rsem(\mathit{input})=v$}
\end{prooftree}
    \caption{Inference rules for CF}
    \label{fig:inferencerules}
\end{figure}

The full inference rule set is given in
Figure~\ref{fig:inferencerules}.
It is essentially the first-order part of Fig. 1 in \cite{jfp}.

\bex The CF program {\tt parity} decides membership in the set $X=\{x\in\{0,1\}^*\  \ | \ \ |x| \mbox{\rm \ is even} \}$.

\bp
     entry x  = even x
     even z   = if (null z) then True else not(even(tail z))
\ep
\eex

\fl This satisfies $\lsem\mbox{\tt parity}\rsem(x) = {\tt True}$ if $|x|$ is even, else {\tt False}.
 For example, $\lsem\mbox{\tt parity}\rsem({\tt[1,0,1]}) = {\tt False}$.
\smallskip

\subsection{The computation tree $ \tpx$ and an evaluation order}
\label{sec-evaluation-step}

\fl{\bf Evaluation steps:} The computation tree $ \tpx$ can be built systematically, applying the rules of Figure~\ref{fig:inferencerules} bottom-up and left-to-right. 
Initially we are given a known program {\tt p} and $\mathit{input} \in \{0,1\}^*$, and the computation goal is $\lsem{\tt p}\rsem(input)=?$, where $?$
is to be filled in with the appropriate value $v$ (if it exists). Intermediate goals are of form ${\tt p},\rho  \vdash {\tt e} \to ?$, where {\tt p}, $\rho$ and {\tt e} are known, and ? is again to be filled in.

Axioms have no premises at all, and base function calls have exactly one premise,
making their tree construction straightforward. The inference rules for conditional expressions have two premises,  and a call to an $m$-ary function has $m+1$ premises.
We choose to evaluate the premises left-to-right. For the case {\tt if e$_0$ then e$_1$ else e$_2$} $\to ?$, there are two possibly applicable inference rules.
However both possibilities begin by evaluating {\tt e$_0$}: finding $v$ such that 
{\tt e$_0$} $\to v$. Once $v$ is known, only one of the two {\tt if} rules  can be applied.

{\em Claim.} the evaluation order above for given {\tt p} and $\mathit{input}$
is general:
if $\lsem{\tt p}\rsem(input)=v$
is deducible by any finite proof at all, then the evaluation order  will terminate with result $\lsem{\tt p}\rsem(input)=v$ at the root.

A consequence:  tree $\tpx$ is unique if it exists, so  CF is a deterministic language. (Nondeterministic variants of CF will considered later.)

\subsection{The suffix property}
\label{sec-suffix-property}

\fl CF programs have no constructors. This implies that all values occurring in the tree $\tpx$ must be boolean values, or suffixes of the input. Expressed more formally:

\begin{definition}
For program {\tt p} and input $x \in \{0,1\}^*$, define its {\em range of  variation}  and its {\em reachable calls} by
\vspace{1mm}

$\ba{lclc}
V_x & = & \{0,1\}\cup \mathit{suffixes}(x) \mbox{\rm\ and} \\
\vspace{1mm}

\mathit{Reach}_{\tt p}(x) & = & \{ ({\tt f},\rho) \ | \  {\tt e}^{\tt f} \mbox{\rm\  is called at least once with environment $\rho$ while computing $\lsem{\tt p}\rsem(x)$}
\}
\ea$
\end{definition}

\begin{lemma}\label{finiteness}
The computation tree  $\tpx$ is finite iff {\tt p} terminates on input $x$.
\end{lemma}

\begin{lemma}\label{suffixes} 

\fl  If  $\tpx$ contains ${\tt p},\rho\vdash {\tt e} \to v$  then $v \in V_x$ and $\rho({\tt z}) \in V_x$ for every ${\tt z} \in \mathit{domain}(\rho)$.
\end{lemma}

Proof: a straightforward induction on the depth of computation trees. If the entry function definition is  {\tt f\,x = e$^{\tt f}$}, then the root $\lsem{\tt p}\rsem(x)=v$ has a parent of the form ${\tt p},\rho_0\vdash {\tt e}^{\tt f} \to v$ where the initial environment is $\rho_0=[{\tt x}\mapsto x]$.
Of course $x$ is a suffix of $x$.
At any non-root node in the computation tree, any new value must either be a Boolean constant or {\tt []}, or constructed  by a base function {\tt not}, {\tt head}, {\tt tail}, or {\tt null} from a value already proven to be in $V_x$.
\vair

\fl Consequence:
{\em a CF computation has at most polynomially many different function call arguments}.

\begin{lemma}\label{polyconfigs}
For any CF program  {\tt p} there is a polynomial $\pi(|x|)$ that bounds $\#\mathit{Reach}_{\tt p}(x)$  for all $x \in \{0,1\}^*$.\end{lemma}
Proof: By Lemma \ref{suffixes}, any argument of any $m$-ary {\tt p}  function {\tt f} has at most $\#V_x=3+|x|$ possible values, so the number of reachable 
argument tuples for  {\tt f} can at most be $(3+|x|)^m$.

{\bf Remark:} this {\em does not necessarily imply} that {\tt p}'s running time is polynomial in $|x|$. We say more on this in Section \ref{timespace}.

\subsection{The tail recursive programming language CFTR}
\label{sec-cftr}

A CFTR program is a CF program  {\tt p} such that every function call 
occurring on the right side of any  defined  function is {\em in tail form}. This is a {\em syntactic} condition; the semantic purpose is to enable a no-stack CFTR implementation, as outlined in Section \ref{sec-lower-level}.
The operational intention is that if a call {\tt f e1...em} is in tail form then, after the call has been performed, control is immediately returned from the current function: there is ``nothing more to do'' after the call.

\bdfn\label{def-alpha}
Define function $\alpha : Exp \to \{X,T,N\}$ as follows, where the set of descriptions is ordered by $X < T < N$.
The intention: if expression {\tt e} contains no function calls then $\alpha({\tt e})=X$,  else if all function calls in {\tt e} are in tail form then
$\alpha({\tt e})=T$, else 
if {\tt e} contains a function call in one or more non-tail-call positions
then $\alpha({\tt e})=N$.
\vair

$\ba{lcll}
\alpha({\tt e}) & = & X & \mbox{if {\tt e} is a constant} \\

\alpha({\tt base\ e}) & = & X & \mbox{if $\alpha({\tt e}) = X$}  \\

\alpha({\tt base\ e}) & = & N & \mbox{if $\alpha({\tt e}) = T$ or  $\alpha({\tt e}) = N$}  \\

\alpha({\tt f\ e1...em}) & = & T & \mbox{if\ } \alpha({\tt e1})=\ldots = \alpha({\tt em})=X\\

\alpha({\tt f\ e1...em}) & = & N & \mbox{otherwise}  \\

\alpha({\tt if\,e0\ then\ e1\ else\ e2}) & = & \max( \alpha({\tt e1}),\alpha({\tt e2})) & \mbox{if\ }\alpha({\tt e0})= X\\

\alpha({\tt if\,e0\ then\ e1\ else\ e2}) & = & N & \mbox{otherwise}  \\
\ea
$

\edfn

\bdfn\label{def-cftr} CF program {\tt p} is in CFTR  iff $\alpha({\tt e})\in\{X,T\}$ for all function definitions {\tt f\ x1...xm = e}.
\edfn

\vair

\fl Some instances in the program {\tt parity}, to clarify the definition:
\be
\item $\alpha({\tt null\ z})=\alpha( {\tt tail\ z})=X$. Neither calls any  program-defined functions.

\item Expression  {\tt even\ x} is a call, and is in tail form: $\alpha({\tt even\ x})=T$.

\item The expression {\tt even(tail\ z)}   is  in tail form: $\alpha({\tt even(tail\ z)})=T$.

\item
\label{item-not-cftr} The expression {\tt not(even(tail\ z))}   is {\em not} in tail form: $\alpha({\tt not(even(tail\ z)}))=N$.

\ee
\vair

\fl By  point \ref{item-not-cftr}, {\tt parity} is {\em not} a CFTR program. However the set $X$ it decides {\em can} be decided by the following alternative  program  whose recursive calls are in tail form. Thus $X$ {\em is} CFTR decidable. 
\vair

\begin{example} Program {\tt parity$'$}
\bp
        entry$'$ x \,=  f x True
        f x y    =  if (null x) then y else f (tail x) (not y)
\ep
\end{example}
\fl A very useful result from \cite{Bhaskar:Simonsen:2020} is the following: we can assume, without loss of generality, that a CF program contains {\em no nested function calls} such as {\tt f x1...xm = ...g(h(e))...}.

\begin{lemma} \label{cf-nesting-removal} 
For any CF program {\tt p} there is a CF program {\tt p}$'$ 
such that \sempar{p} = \sempar{p$'$}, and for  each function definition {$\tt f\ x1\ ...\ xm = e^f$} in {\tt p}$'$, either $\alpha({\tt e^f}) \leq T$, or ${\tt e^f} = {\tt if\ e_0\ then\ e_1\ else\ e_2}$
and $\alpha({\tt e_i}) \leq$ T for $i \in \{0,1,2\}$. 
\end{lemma}

\fl 

An interesting step in the proof of Lemma \ref{cf-nesting-removal} is to show that a  function $f : X \to Y$ is CF-computable if and only if its graph 
$G^f = \{(x,y) \in X\times Y \ | \ f(x)=y\}$ is CF-decidable (see \cite{Bhaskar:Simonsen:2020} for details).
In a CFTR program, no function calls are allowed to occur in the test part of a  tail-recursive
conditional expression (by the last line of the $\alpha$ definition). Thus 
Lemma \ref{cf-nesting-removal}  does not rule out a call nested in a conditional's test part, and so does not imply that {\tt p$'$} is tail-recursive.

An example that suggests that {\em some} call nesting is essential: the $MCV$ program of Section \ref{sec-mcv} has function calls inside  {\tt if} tests. It is well-known 
that the function that $MCV$ computes can be computed tail-recursively if and only if {\sc logspace}={\sc ptime} \cite{papadimitriou}.

\section{About time and space measures}\label{timespace}

Theorems 1 and 2 relate computations by two rather different computation models:  cons-free programs and Turing machines. We now focus on  time and space relations between the two. We define the time and space used to run given CF program {\tt p} on input $x \in \{0,1\}^*$:

\bdfn\hfill

\bi

\item 
 ${\mathit time}_{\tt p}(x)$ is the number of evaluation steps to run program {\tt p} on $x$ (as in Section \ref{sec-evaluation-step}).  

\item
${\mathit space}_{\tt p}(x)$ is the number of bits required to run  {\tt p} on input $x$.  
\ei
\edfn

\fl Both ${\mathit time}_{\tt p}(x)$ and ${\mathit space}_{\tt p}(x)$ are non-negative integers (or $\bot$  for nontermination). 
We call ${\mathit time}_{\tt p}(x)$ the ``{\em native time}" for CF or CFTR (in contrast to Turing machine time). 
For the  example, ${\mathit time}_{\tt parity} (x) = O(|x|)$.
It is reasonably simple to define running time as
${\mathit time}_{\tt p}(x) = |\tpx|$, i.e., the number of nodes in the computation tree  $\tpx$.

A full definition of ${\mathit space}_{\tt p}(x)$   requires a lower-level execution model than the abstract semantics above. (Section \ref{sec-lower-level} will show a bit more detail.) 
Consider the {\tt parity} example.
The length of the call stack needed to implement the program is linear in the depth of the proof tree, and so $O(|x|)$.
Moreover, each non-root node has an environment $\rho$ that binds {\tt x} or {\tt z} to its value.
By Lemma \ref{suffixes} any such value is a suffix of the input.
Any suffix can be represented in $O(\log |x|)$ bits, so the total space that {\tt parity} uses is
$${\mathit space}_{\tt parity}(x) = O(|x|\log |x|)$$

\subsection{Space usage: a  lower-level operational semantics, and the tail call optimisation}  
\label{sec-lower-level}

To more clearly define space usage and present the tail call optimisation, we use a finer, more operational level of detail based on traditional implementations\footnote{Readers may think of machine-code run-time states, e.g.,  value stacks, stack frames,\ldots; or of tapes and state transition relations. Such implementation paraphernalia are well-known  both to compiler writers and programming language theorists.}.
Programs are executed sequentially as in the proof-tree-building algorithm seen earlier.   The expression on the right side of a function definition is evaluated one step at a time as in Section \ref{sec-evaluation-step}; but with an extra data structure, a stack $\sigma$ used to record some of the environments  seen before. The stack's topmost activation record corresponds closely to the current environment $\rho$.

Each defined function {\tt f x1...xm = e} has an activation record format that contains the values $v_1,\ldots,v_m$ of the arguments with which {\tt f} has been called (plus shorter-lived ``temporary results,'' i.e., as-yet-unused subexpression values).  On any call to {\tt f}, a new activation record is pushed onto the stack $\sigma$. This activation record will be popped when control returns from the call.

\begin{definition}\label{reachableconfiguration} The {\em call history} for CF program {\tt p} and  input $x \in \{0,1\}^*$ is a  
sequence (finite or infinite) of   {\em  configurations}, each of form 
$({\tt f},\rho) \in \mathit{Reach}_{\tt p}(x)$ where {\tt f} is a defined function name  and $\rho$ is an environment.
The first configuration always has form $({\tt p-entry},[{\tt z} \mapsto x[)$.
 
\end{definition}

\fl How  the syntactic CFTR condition in
Definition \ref{def-cftr} allows more efficient implementation:

\subsubsection*{The tail call optimisation}
Claim: every CFTR program can be implemented {\em using at most one  stack frame}
at a time.

Suppose CF program {\tt p} has a definition {\tt f x1...xr = ...(g e1...es)...} where the {\tt g} call is a tail call.
If so, then after the call {\tt (g e1...es)} has been started there can be no future references to the current values of {\tt x1,...,xr}.
Informally, there is ``nothing more to do'' after a tail call has been completed. (One can review the {\tt parity'} and {\tt parity} examples in this light.)

The   tail call optimisation: The new activation record for {\tt g}  {\em overwrites}  {\tt f}'s current activation record.
In a CFTR program {\em every} call will be a tail call. 
Thus there is never more than one frame at a time in the runtime stack. 
One stack frame requires $O(\log|x|)$ bits. For example, this
gives {\tt parity}$'$ a major improvement over the space used by {\tt parity}: $
    {\mathit space}_{{\tt parity}'}(x) = O(\log |x|)
    $ bits.
\vair

\fl{\bf A relaxation of the tail call restriction.}
Sketch (details in \cite{Bhaskar:Simonsen:2020}): It is sufficient that the functions defined in {\tt p} can be given a partial order $<$ such that no call is to another that is earlier in the order, and recursive calls must be  tail calls. 
Under this lighter restriction there may be more than one frame at a time in the runtime stack; but these frames are properly ordered by $<$.
Consequently every restricted program can be implemented within a {\em constant depth of  stack frames}.

\subsection{CFTR programs run in polynomial time; but CF 
can take exponential time}\label{subsec:nonpolynomial-run}

\begin{theorem}\label{cftrpolytime}Any terminating CFTR program runs in (native) time polynomial in $|x|$.
\end{theorem}
Proof: in the call history of {\tt p} on input $x$, all runtime configurations must be distinct (else the program is in an infinite loop).
There is a constant bound on the number of stack frames.
Thus, by Lemma \ref{polyconfigs} there are only polynomially many distinct runtime configurations.

\begin{theorem}\label{cfexptime} A terminating run of a CF program can take   time exponential in its input length.
\end{theorem}
First, a trivial example with non-polynomial running time (from \cite{jfp}):
\bex
\bp 
   f x = if (null x)  then True  else 
         if f(tail x) then f(tail x) else False      
\ep
\eex
It is easy to see that $\mathit{time}_{\tt q}(x) = 2^{O(|x|)}$ due to the repeated calls to {\tt tail x}.
Exponential time is not necessary, though, since the computed function  satisfies  {\tt f(x) = True}.

\subsection{A more fundamental example: a CF-decidable problem that is {\sc ptime}-complete }
\label{sec-mcv}

$MCV$ is the \emph{Monotone Circuit Value Problem}. Technically it is the set of all {\em straight-line Boolean programs} whose last computed value is {\tt True}, e.g.,
\smallskip

$x_0 := {\tt False}; \ x_1 := {\tt True}; \ x_2 := x_1 \lor x_0; \ x_3 := x_2 \land x_0; \ x_4 := x_3 \lor x_2; \ x_5 := x_4 \lor x_3
$
\smallskip

\fl $MCV$ is a well-known ``hardest'' problem for {\sc ptime}
\cite{papadimitriou}. In complexity theory, completeness means two things: (1) $MCV \in \mbox{\sc ptime}$; and (2) if $X \subseteq \{0,1\}^*$ is {\em any} problem in {\sc ptime}, then $X$ is 
logspace-reducible\footnote{Details may be found in \cite{Jones:97:ComputabilityComplexity}, Chapters 25, 26.}  to $MCV$. 
By transitivity of logspace-reduction, $MCV \in \mbox{\sc logspace}$ if and only if $ \mbox{\sc logspace} = \mbox{\sc ptime}$.

Following is the core of a Haskell program to decide membership in $MCV$, together with a sample run.
We chose a compact Boolean program representation, so the program above becomes a list:
\bp 
 program = [ OR 5 4 3, OR 4 3 2, AND 3 2 0, OR 2 1 0 ]
 \ep
{\bf Haskell encoding:} Assignment $x_i := x_j \land x_k$ is represented by omitting $:=$, and coding $\land$ as {\tt AND} in prefix position (and similarly for $\lor$). The program is presented in reverse order; and variables {\tt X0, X1} always have predefined values {\tt False, True} resp.,  so their assignments are omitted in the representation.

\bp
type Program     =  [Instruction]
data Instruction =  AND Int Int Int  |  OR Int Int Int

mcv :: Program       -> Bool
vv ::  (Int,Program) -> Bool

mcv ((OR  lhs x y):s) = if vv(x,s) then True    else vv(y,s)
mcv ((AND lhs x y):s) = if vv(x,s) then vv(y,s) else False
             
vv(0,s) = False --  vv(v,s)  =  value of variable v at program suffix s
vv(1,s) = True  --              
vv(v,s) = case s of 
           ((AND lhs x y):s') -> if v==lhs then mcv(s) else vv(v,s')
           ((OR  lhs x y):s') -> if v==lhs then mcv(s) else vv(v,s')
           
\ep
This works by recursive descent, beginning at the end of the program; and uses no storage at all (beyond the implicit implementation stack). 
 Following is a trace of nontrivial calls  to {\tt vv}, ended by the final result of running {\tt p}. Note that variable {\tt X2} is evaluated repeatedly.
\bp
   program = [X5:=X4 OR X3, X4:=X3 OR X2, X3:=X2 AND X0, X2:=X1 OR X0]

vv(X4,[ X4 := X3 OR  X2, X3 := X2 AND X0, X2 := X1 OR  X0])
vv(X3,[ X3 := X2 AND X0, X2 := X1 OR  X0])
vv(X2,[ X2 := X1 OR  X0])
vv(X2,[ X3 := X2 AND X0, X2 := X1 OR  X0])
vv(X2,[ X2 := X1 OR  X0])

True

\ep

\fl {\bf Re-expressing the Boolean program as a bit string for CF input}
\vair

\fl Of course CF  programs do not have data values of type {\tt Int}. However, any straight-line Boolean program can be coded by a CF bit string. Example: the  Boolean program above can be coded as a 44-bit string\footnote{The {\tt --} parts delimit comments, and are not part of the Boolean program's bit-string encoding.}:
\bp
[1,1,1,0,   1,0,1, 0, 1,0,0, 0,1,1,                 -- x5 := OR  x4  x3
            1,0,0, 0, 0,1,1, 0,1,0,                 -- x4 := OR  x3  x2
            0,1,1, 1, 0,1,0, 0,0,0,                 -- x3 := AND x2  x0 
            0,1,0, 0, 0,0,1, 0,0,0]                 -- x2 := OR  x1  x0
            
\ep
Bag of tricks: 
The program has 5 variables. The index {\tt i} of any  variable {\tt xi} can 
be coded by a $3$-bit binary sequence we denote by $\hat{\tt i}$, e.g., {\tt x3}  is coded by 
$\hat{\tt 3}$ = {\tt 0,1,1}. This has length $3$ since $3=\lceil\log_2 5\rceil$. 

The beginning 
{\tt 1,1,1,0} of the program 
code gives (in unary) the code block length, $3$ for this program. 
A Boolean assignment 
{\tt xi := xj op xk} is coded as $\hat{\tt i}\ \hat{\tt op}\ \hat{\tt j}\ \hat{\tt k}$, where the Boolean operators $\lor,\land$ are coded as {\tt 0}, {\tt 1} respectively. 

Using this encoding, it is not hard to transform the Haskell-like program to use if-then-else statements rather than case; and then implement it in the true CF language.
 
\vair

\fl{\bf A dramatic contrast in running times}\hfill
\bi
\item The CF program just sketched has repeated (and nested) function calls with the same argument values. For example, Boolean variable {\tt X2} in program {\tt p}
is evaluated 3 times in the trace above. More generally, the running time bound is 
exponential:
$$\mathit{time}_{\tt mcv}(x) = 2^{O(|x|)}
$$

\item On the other hand, one can show that $MCV$ {\em can be decided by a Turing machine} in
polynomial time: Execute the Boolean instructions from the beginning,  store the values of left side variables when computed, and refer to stored variable values as needed.
\ei
This major difference is due to data storage: the Turing machine tape can save already-computed values, to be looked  up as needed. A CF program has no such storage, and must resort to recomputation\footnote{Duplicating CF variables does not suffice, since the number of variables is independent of the length of the  input data.}. 
\smallskip

\fl{\bf Is this contrast necessary?} It seems strange that the cost of strengthening tail recursive programs (in CFTR) by adding recursion (to get CF) is to raise run times from polynomial to exponential.
The next sections, abbreviated from \cite{jfp}, show
that the problem is a general one of the relation between {\sc logspace} and {\sc ptime}. Thus it is not peculiar to the MCV problem, nor to the way we have programmed its solution.
         
\section{How $\mbox{\sc logspace} = \decidepar{CFTR}$ and 
$ \mbox{\sc ptime} =  \decidepar{CF}$ are proven} 
\label{howmainresultsproved}

\subsection{Proof sketch of Theorem \ref{thm-l-equivalence}, that $\mbox{\sc logspace} = \decidepar{CFTR}$}

\bi

\item  For $\supseteq$, we simulate a CFTR program {\tt p} by a {\sc logspace} Turing machine (informal).
Given an input $x=a_1\ldots a_n \in\{0,1\}^*$, by Lemma \ref{suffixes} any reachable configuration $({\tt f},\rho) $
satisfies $0 \leq |\rho({\tt xi})| \leq \max(n,1)$ for $i=1,\ldots,m$.
Each $v_i$ can be coded in $O(\log n)$ bits. Now {\tt f} and the number of {\tt f}'s arguments are independent of $x$, so
an entire configuration $({\tt f},\rho) $ can be coded into $O(\log n)$ bits.

The remaining task is to show that the operational semantics of running {\tt p} on $x$ can be simulated
by a {\sc logspace} Turing machine. The key: because {\tt p}
is  tail recursive, there is {\em no nesting} of calls to program-defined functions.
The construction can be described by cases:
\be
\item Expressions without calls to any defined function: Suppose {\tt p}'s current environment $\rho$
is represented on the Turing machine work tape. Simulating this evaluation is straighforward. 

\item Evaluation of an expression {\tt f\,e1\ldots em}: Since {\tt p} is tail recursive, none of {\tt e1,...,em} contains a call, and there is ``nothing more to do'' after the call {\tt f\,e1\ldots em} has been simulated. Assume inductively that {\tt e1\ldots em} have all been evaluated. Given the definition {\tt f\,x1\ldots xm = e} of {\tt f}, the remaining 
steps\footnote{These steps are done using Turing machine representations of the environments as  coded on  Tape 2.} 
are to collect the values $v_1,\ldots,v_m$ construct a new environment 
$\rho' = [{\tt x1}\mapsto v_1,\ldots,{\tt xm}\mapsto v_m]$, and replace the current $\rho$ by $\rho'$.

\ee

\vair

\item  For $\subseteq$, we simulate a {\sc logspace} Turing machine by a CFTR program. 
This can be done using Lemma \ref{suffixes}:  represent a Tape 2 value  by one or more suffixes of  $x$. (A more general result is shown in \cite{jfp}  by  ``counting modules''.)


\ei

\fl {\bf Remark:} given an input $x$, the 
{\em number of times} that a call $({\tt f},\rho)$ appears in the call history of $\lsem{\tt p}\rsem(x)$ may be much larger than the number of such calls {\em with distinct argument values}, even exponentially larger.

\subsection{Proof sketch of  Theorem \ref{thm-p-equivalence}, that $\mbox{\sc ptime} =  \decidepar{CF} $}
\label{sec-timeequalsCF}

\bi
\item For $ \subseteq$, we simulate a {\sc ptime} Turing machine $Z$ by a CF program.
(this is the core of {\em Proposition} 6.5 from \cite{jfp}). Assume Turing machine
$Z=(Q,\{0,1\},\delta, q_0)$ is given\footnote{As usual $Q$ is a finite set with initial state $q_0$, and transition function of type
 $\delta: Q \times \{0,1,B\} \to Q \times \{0,1,B\}  \times \{0,1,-1\}$.
 A {\em total state} is a triple $(q,i,\tau)$ where $q\in Q, i\geq 0$ and the tape is $\tau: \nats \to  \{0,1,B\}$. 
 Scan positions are counted $i = 0,1,\ldots$ from the tape left  end.
 Transition
 $\delta(q,a) = (q',a',d)$ means: if $Z$ is in state $q$ and $a = \tau(i)$ is the scanned tape symbol, then change the state to $q'$, write $a'$ in place of the scanned symbol, and move the read head from its current position $i$ to position $i+d$.} and that it
 runs in time at most $n^k$ for inputs $x$ of length 
 $n$.
Consider an input
$x = \pgt{a}_1\pgt{a}_2\ldots\pgt{a}_n\in\{\pgt{0},\pgt{1}\}^*$.

Idea: in the space-time diagram of $Z$'s computation on $x$
 (e.g., Fig. 3 in \cite{jfp}), data propagation 
is local; the information (control state $q$,  symbol $a$, whether or not it is scanned) at time
$t$ and tape position $i$ is a mathematical function of the same
information at time  $t-1$ and tape positions $i-1, i, i+1$. This connection
can be used to compute the contents of any tape cell at any time. Repeating for $t=0,1,\ldots,n^k$ steps gives the final result of the computation ($x$ is accepted or not accepted). 

The simulating CF program computes functions\vair

\hspace{10mm}\bt{lcl}

$\pgt{state}(t)$  &= &  the state $q$ that $Z$ is in 
at time
$t$ \\

$\pgt{position}(t)$ &= & the
scanning position $i\in\{0,1,2,\ldots,n^k\}$ at
time
$t$\\

$\pgt{symbol}(t,i)$  &= &  the tape symbol $a$ found at time
$t$ and scanning position  $i$\\ \\

\et

\fl Initially, at time $t=0$ we have total state
$(q_0,0,\tau_0)$ where $\tau_0(i) = {\tt a}_i$ for $1 \leq i \leq n$, else $\tau_0(i) = B$.
Now suppose $t>0$ and the total state at time $t-1$ is 	$(q,i,\tau)$, and that $\delta(q,a) = (q',a',d)$.
Then the total state at time $t$ will be
$(q',i+d,\tau')$ where $\tau'(i) = a'$ and $\tau'(j)=\tau(j)$ if $j\neq i$. 

\vair

Construction of a CF program {\tt z} with definitions of \pgt{state}, \pgt{position} and \pgt{symbol}
is straightforward. Arguments $t, i\in \{0,1,2,\ldots,n^k\}$ can be uniquely encoded as tuples of suffixes of 
$x = \pgt{a}_1\pgt{a}_2\ldots\pgt{a}_n\in\{\pgt{0},\pgt{1}\}^*$.
It is not hard to see that such an encoding is possible; \cite{jfp} works out  details for an imperative version of the Turing machine (cf.\ Fig. 5).

\item For $\supseteq$, we
 simulate an arbitrary CF program {\tt p} by a {\sc ptime} Turing machine (call it $Z_{\tt p}$). 
 We describe the $Z_{\tt p}$ computation informally.
Given an input $x\in\{0,1\}^*$, $Z_{\tt p}$ will systematically build a {\em cache} containing $\mathit{Reach}_{\tt p}(x)$. By Lemma \ref{polyconfigs} its size is polynomially bounded.

The cache (call it $c$) at any time contains a set of triples $({\tt f},\rho, v')$ where
the simulator has completed evaluation of function {\tt f} on argument values in environment $\rho$, and  this {\tt f}
call has returned value $v$.

Concretely, $c$ is built ``breadth-first'': when a {\tt p} call {\tt f\,e1\ldots em} is encountered (initially $c$ is empty):
\bi
\item First, arguments  {\tt e1,\ldots,em} are evaluated to values $v_1,\ldots,v_m$; and these are collected into an environment $\rho$.
\item Second, cache $c$ is searched. If it contains $({\tt f},\rho, v)$, then simulation continues using the value $v$ as the value of {\tt f\,e1\ldots em}.
\item If  $c$ contains no such triple,  then {\tt p} must contain a function definition {\tt f\,x1...xm = e$^{\tt f}$}.
Then expression {\tt e}$^{\tt f}$ is evaluated in environment $\rho$ to yield some value $v$. After finishing, add the triple $({\tt f},\rho, v)$ to $c$, and return the value $v$.
\ei

\ei

\fl{\bf An observation:}
{\em The CF program of Section \ref{sec-timeequalsCF} (simulating a polynomial time Turing machine) has exponential runtime.}
\vair

\fl
Paradoxically, as  observed in Section \ref{subsec:nonpolynomial-run}: even though CF exactly characterises {\sc ptime}, its programs do not run in polynomial time.
The polynomial versus exponential  contrast between the running times of CFTR and CF programs is  interesting since both program classes are natural; and the {\em decision powers} of CFTR and CF are exactly the complexity classes  {\sc logspace} and {\sc ptime}. Alas, we have found no  CF algorithm to simulate polynomial-time Turing machines.
We  explain how this happens in more detail.


\begin{definition}
{\em Call overlap} occurs if a CF program can call the same function more than once with the same tuple of argument values.
\end{definition}

\fl{\bf How can this happen?} By Lemma \ref{polyconfigs}, only polynomially many argument tuples are distinct.
Consequently, superpolynomial running time implies that some simulation functions may be called repeatedly with the same argument value tuples, 
even though only polynomially many of them  are distinct.

This can be seen in the proof that $ \mbox{\sc ptime} \supseteq \decidepar{CF}$: the value of  $\pgt{symbol}(t+1,i)$ may depend on the values of 
$\pgt{symbol}(t,i-1)$, and $\pgt{symbol}(t,i)$, and  $\pgt{symbol}(t,i+1)$.
In turn,  the value of 	$\pgt{symbol}(t,i)$ may depend on the values of 
$\pgt{symbol}(t-1,i-1)$, and $\pgt{symbol}(t-1,i)$, and $\pgt{symbol}(t-1,i+1)$.

Net effect: a symbol {\tt a}$_i$ on the final tape (with $t=n^k$) may depend many distinct function call sequences
that ``bottom out'' at $t=0$.
The number of call sequences may be exponential in $n$.

\vair

\begin{lemma}
Call overlap will occur if the length of the call history for $\sempar{p}(x)$
is greater than $\# Reach_{\tt p}(x) $.
\end{lemma}
Unfortunately, it is  is hard to see {\em which} calls will overlap (it seems impossible without storage).
Furthermore, the ``caching'' technique used to prove Theorem \ref{thm-p-equivalence}
cannot be used because, in contrast with Turing machines, CF programs do not have a memory that can accumulate
previously computed values.

\section{Closer to the boundary between CFTR and CF}

Viewed {\em extensionally}, the difference (if any) between CFTR and CF corresponds to the difference  (if any) between {\sc logspace} and {\sc ptime}. Viewed {\em intensionally}, there seem to be significant differences, e.g., in algorithm expressivity as well as in running time. A relevant early result:
\vair

\fl{\bf A program transformation by Greibach} shows that  programs whose  
calls (to defined functions) are all linear\footnote{A  call is linear if it is not contained in a call to any defined function. An example with a linear call that is not a tail call: {\tt not(even(tail z))} in Section \ref{sec-cftr}.}
can be made tail recursive \cite{DBLP:books/sp/Greibach75}.
Using this transformation, a CF program {\tt p} whose calls are tail calls or linear calls may be transformed into one containing
only tail calls (and no CONS or other constructors), so it is in CFTR. Thus {\tt p}  decides a problem in {\sc logspace}.

There is a price to be paid, though: in general, the transformed program may run {\em polynomially slower than the original}, due to re-computation. For instance, the Greibach method can transform our first ``parity'' program into CFTR form. The cost: to raise time complexity from linear to quadratic. 
\vair{}

\fl Following is a  new tool to investigate the problem. It is analogous to the set of programs for polynomial-time Turing machines, but adapted to the CF world. By Theorem \ref{cfexptime}, CFpoly $\subsetneq$ CF.
\vair{}

\bdfn

The programming language CFpoly has the same semantics as CF; but CFpoly's programs are restricted to be those CF programs {\em that terminate in polynomial time}.
\edfn 
Immediate: 
$$\mbox{\sc logspace} = \decidepar{CFTR} \subseteq  \decidepar{CFpoly}  \subseteq  \decidepar{CF}\ = \mbox{\sc ptime}
$$
By Theorem \ref{cftrpolytime}, every terminating CFTR program is in CFpoly. 
One can see Greibach's result as transforming a subset of CFpoly into CFTR.

\vair{}

\begin{lemma}
If a terminating CF program does not have call overlap, then it is in CFpoly.
\end{lemma}

The reason: by Lemma \ref{polyconfigs}, such a CF program
must run in polynomial time. 
\vair

\fl{\bf Remark:} On the other hand, it may have  non-tail calls, and so not be in CFTR.
\vair

\begin{lemma}
Problem class $\decidepar{CFpoly}$  is closed under $\cup,\cap$ and complement.
\end{lemma}

\section{Nondeterminism and cons-free programs}
\label{sec-nondeterminism}

Nondeterministic programs allow expression evaluation and program runs to be relations ($\leadsto$) rather than functions. 
A nondeterministic version of
Figure \ref{fig:inferencerules}
would use 
 $\lsem{\tt p}\rsem(x) \leadsto v$ and ${\tt p}, \rho \vdash {\tt e} \leadsto v$
 in place of $\lsem{\tt p}\rsem(x) \to v$  and ${\tt p}, \rho \vdash {\tt e} \to v$.
 Alas, the algorithm of Section \ref{sec-evaluation-step} cannot be used if {\tt p} is nondeterministic.

\begin{definition}
A set $X \subseteq  \{0,1\}^*$ is {\em decided} by an NCF program {\tt p} if
for all inputs $x \in \{0,1\}^*$, 
$$
x\in X \mbox{\ if and only if {\tt p} has a computation\ } \lsem{\tt p}\rsem(x) \leadsto {\tt True}$$
\end{definition}

\subsection{Nondeterministic versions of the CF program classes}

\fl{\bf Remark:} the reasoning used in Theorem \ref{howmainresultsproved} clearly extends to show that $\mbox{\sc nlogspace} = \decidepar{NCFTR}$.
The following is particularly interesting since the question {\sc logspace} =? {\sc nlogspace} is still open.

\begin{theorem}
$\decidepar{CF} = \decidepar{NCF}$
\end{theorem}
\vair{}

\fl This was proven by Bonfante \cite{DBLP:conf/amast/Bonfante06} by a technique that stems back to in Cook \cite{Cook:71:CharacterizationOfPushdown}. 
The implication is that $\decidepar{CF}$ is closed under nondeterminism, since
both problem classes are equal to {\sc ptime}. 
(This   does {\em not} imply {\sc ptime} = {\sc nptime},
since it is not hard to see that any NCF program can be simulated by a deterministic polynomial-time Turing machine: Lemma \ref{polyconfigs}  holds for NCF as well as for CF.)
\vair

However it is  not known whether  $\decidepar{CFpoly}$ is closed under nondeterminism, since Bonfante and Cook's reasoning does not seem to apply. Why? The memoisation used in \cite{DBLP:conf/amast/Bonfante06} yields  a 
{\em Turing machine} polynomial time algorithm. Consequently,  the problem is decidable by some CF program; but we know no way  to reduce its time usage  from exponential  to polynomial.
\vair{}

\fl {\bf A ``devil's advocate'' remark:} by Theorem \ref{thm-p-equivalence} the
 classes $\decidepar{CFpoly}$ and {\sc nlogspace} are both between \mbox{\sc logspace} and \mbox{\sc ptime}:
$$
\mbox{\sc logspace} = \decidepar{CFTR} \subseteq \mbox{\sc nlogspace}    \subseteq  \decidepar{CF} = \mbox{\sc ptime}
$$
So why bother with yet another class?
One answer: 
CFpoly seems more natural than NCFTR from a programming viewpoint; and
intuitively $\decidepar{CFpoly}$ seems to be a larger  extension of  {\sc logspace} than {\sc nlogspace} or the program class NCFTR. 
However we have no proof that $\mbox{\sc nlogspace} \neq \decidepar{CFpoly}$, and no solid reason to think that either class contains the other.

\subsection{Simulating CFpoly by a nondeterministic algorithm}

\begin{theorem}\label{thm-cynthias-result}
$\decidepar{CFPoly} \subseteq$ NSPACE(log$^2$ n)).
\end{theorem}

\fl{Proof:}
\fl If \texttt{p} $\in$ CFPoly, there is a polynomial $\pi$ such that for any input $x \in \{0,1\}^*$ one can decide, in time $\pi(|x|)$, 
whether or not $\lsem{\tt p}\rsem(x) \to v$. The question: can this be done in significantly less space? We answer ``yes''.
The proof uses a space-economical nondeterministic algorithm to find $v$ at the root $\lsem{\tt p}\rsem(x)=v$ of  tree $\tpx$. First, observe that any reachable statement 
$\mathtt{p},\rho \vdash {\tt e} \rightarrow w$ can be represented in
$O(\log n)$ bits.
\begin{itemize}
\item To evaluate \texttt{p} on input $x$, we nondeterministically \emph{guess} a value  $v$ such that
  $\lsem{\tt p}\rsem(x) = v$.  The remaining task is to \emph{confirm} that this statement is true.  If we cannot do that, the whole algorithm has failed.
\item To \emph{confirm} a statement such as $\mathtt{p},\rho \vdash {\tt f\ e} \to v$, 
  we must confirm the existence of an evaluation tree of the form:
  \begin{prooftree}
  \AxiomC{\dots}
  \UnaryInfC{$\mathtt{p},\rho \vdash {\tt e} \rightarrow w$}
  \AxiomC{\dots}
  \UnaryInfC{$\mathtt{p},[\bind{x_1}{w}] \vdash {\tt e^f} \rightarrow v$}
  \BinaryInfC{$\mathtt{p},\rho \vdash {\tt f\ e} \rightarrow v$}
  \end{prooftree}
  To do this, we now \emph{guess} $w \in V_x$, and now need to confirm two
  statements.  We also \emph{guess} which of these two statements has the {\bf shortest}
  evaluation tree.
  
  For example, suppose we guess that this is the case for $\mathtt{p},
  [\bind{x_1}{w}] \vdash {\tt e^f} \rightarrow v$.  Then we do a recursive call to confirm
  this statement (which temporarily stores the current state on the 
  algorithm's stack).
  Afterwards, we {\em tail-recursively} confirm the other statement, $\mathtt{p},\rho \vdash {\tt e} \rightarrow w$. Since this is a tail call, the algorithm's stack size does not increase.
  
\item This extends naturally to multi-argument function calls. 

\item To confirm a statement $\mathtt{p},\rho \vdash \mathtt{if}\ e_1\ 
  \mathtt{then}\ e_2\ \mathtt{else}\ e_3 \to v$, we \emph{guess} whether $e_1$
  reduces to True or False; for example, we guess that it reduces to False.  Then
  it suffices to confirm that an evaluation tree of the following form exists:
  \begin{prooftree}
  \AxiomC{\dots}
  \UnaryInfC{$\mathtt{p},\rho \vdash e_1 \rightarrow $False}
  \AxiomC{\dots}
  \UnaryInfC{$\mathtt{p},\rho \vdash e_3 \rightarrow v$}
  \BinaryInfC{$\mathtt{p},\rho \vdash \mathtt{if}\ e_1\ \mathtt{then}\ e_2\ \mathtt{else}\ e_3 \rightarrow v$}
  \end{prooftree}
  For this, we again have to confirm two statements. As before, we \emph{guess}
  which of the two statements has the shorter evaluation tree, evaluate that
  statement first, and then evaluate the other statement tail-recursively.
\end{itemize}
If all guessed values are correct, then this algorithm returns the correct result.
Furthermore, if we always guessed correctly which statement has the shortest evaluation tree, it does so with an overall stack depth that is never more than log($|\mathcal{T}^{\mathtt{p},x}|$).
The reason is that every time we do a non-tail recursive call,  this shortest subtree has size at most half of the previous subtree's size.  Since a statement 
$\mathtt{p},\rho \vdash {\tt e} \rightarrow w$ can be represented in
$O(\log n)$ bits we have the desired result.

The idea of applying tail-recursion to the largest subtree to save space
was also used to implement Quicksort~\cite{Knuth:73:ArtOfProgramming}.
However our problem is much more general; and we must use nondeterminism because it cannot be known in advance which subtree will be the largest.

\vair

\fl We expect that this construction can be extended to show $\decidepar{NCFpoly} \subseteq \mbox{\sc nspace}(\log^2 n)$ as well.

\subsection{Closure under complementation}
The following  was shown by a subtle nondeterministic state-enumeration and state-counting algorithm.
It was devised independently by Immerman and Szelepcs{\'{e}}nyi, and solved Kuroda's second and long-standing open question \cite{Kuroda,Immerman,Szelepcsenyi}. 
It is still open whether $\mbox{\sc logspace} \subsetneq 
\mbox{\sc nlogspace}$.

\begin{theorem}
{\sc nlogspace} is closed under complement. 
\end{theorem}
\label{thm-nspace-complement}

\fl

\section{Conclusions, future work}

We have probed the question {\sc logspace} =? {\sc ptime} 
from a programming language viewpoint. 
A starting point was that the ``cons-free'' programming language CF exactly captures the Turing complexity class {\sc ptime}; while its
cons-free tail recursive subset CFTR exactly captures {\sc logspace}.
Section \ref{howmainresultsproved} recapitulates the reasoning used in \cite{jfp}.

In more detail: all CFTR programs run in polynomial time; but on the other hand, some CF programs run for exponentially many steps. Further the sets decided by the two program classes have seemingly different closure properties: the questions {\sc logspace} =? {\sc nlogspace} and
{\sc ptime} =? {\sc nptime} and even 
{\sc logspace} =? {\sc ptime}
have been open for decades.
Given this, it seems almost paradoxical that $\decidepar{CF} = \decidepar{NCF}$ (from \cite{DBLP:conf/amast/Bonfante06}).

Trying to understand these differences made it natural to consider CFpoly - the class of polynomially time-bounded CF programs since they have feasible running times (even though some CF programs may have superpolynomial behavior=.
One test of  CFpoly was to see whether it contained any 
{\sc ptime}-complete problems. As a case study we wrote (Section \ref{sec-mcv})
a CF-program for  MCV (the monotone circuit value) problem
to clarify where non-tail-recursion was necessary, and where superpolynomial runtimes came into the picture. (One key was nested recursion in function calls, clearly visible in function {\tt mcv} in the MCV code.)
Another test was to see whether 
$\decidepar{CFpoly}$ is perhaps a smaller complexity class than {\sc ptime}.
Theorem \ref{thm-cynthias-result} leads in this direction, with an interesting proof construction and the surprising upper bound $NSPACE(\mathrm{log}^2 n)$.

Many questions for CFpoly are still to be  investigated. One is to see whether the Immerman-Szelepcsenyi
algorithm can be adapted to NCFpoly.

\bibliographystyle{eptcs}

\bibliography{main}

\begin{thebibliography}{10}
\providecommand{\bibitemdeclare}[2]{}
\providecommand{\surnamestart}{}
\providecommand{\surnameend}{}
\providecommand{\urlprefix}{Available at }
\providecommand{\url}[1]{\texttt{#1}}
\providecommand{\href}[2]{\texttt{#2}}
\providecommand{\urlalt}[2]{\href{#1}{#2}}
\providecommand{\doi}[1]{doi:\urlalt{http://dx.doi.org/#1}{#1}}
\providecommand{\bibinfo}[2]{#2}

\bibitemdeclare{techreport}{Bhaskar:Simonsen:2020}
\bibitem{Bhaskar:Simonsen:2020}
\bibinfo{author}{Siddharth \surnamestart Bhaskar\surnameend} \&
  \bibinfo{author}{Jakob~G. \surnamestart Simonsen\surnameend}
  (\bibinfo{year}{2020}): \emph{\bibinfo{title}{Implicit Complexity via
  Controlled Construction and Destruction}}.
\newblock \bibinfo{type}{Technical Report}, \bibinfo{institution}{Department of
  Computer Science, University of Copenhagen}.

\bibitemdeclare{inproceedings}{DBLP:conf/amast/Bonfante06}
\bibitem{DBLP:conf/amast/Bonfante06}
\bibinfo{author}{Guillaume \surnamestart Bonfante\surnameend}
  (\bibinfo{year}{2006}): \emph{\bibinfo{title}{Some Programming Languages for
  Logspace and Ptime}}.
\newblock In \bibinfo{editor}{Michael \surnamestart Johnson\surnameend} \&
  \bibinfo{editor}{Varmo \surnamestart Vene\surnameend}, editors: {\sl
  \bibinfo{booktitle}{Algebraic Methodology and Software Technology, 11th
  International Conference, {AMAST} 2006, Kuressaare, Estonia, July 5-8, 2006,
  Proceedings}}, {\sl \bibinfo{series}{Lecture Notes in Computer Science}}
  \bibinfo{volume}{4019}, \bibinfo{publisher}{Springer}, pp.
  \bibinfo{pages}{66--80}, \doi{10.1007/11784180\_8}.

\bibitemdeclare{article}{Cook:71:CharacterizationOfPushdown}
\bibitem{Cook:71:CharacterizationOfPushdown}
\bibinfo{author}{Stephen~A. \surnamestart Cook\surnameend}
  (\bibinfo{year}{1971}): \emph{\bibinfo{title}{Characterizations of Pushdown
  Machines in Terms of Time-Bounded Computers}}.
\newblock {\sl \bibinfo{journal}{J. {ACM}}}
  \bibinfo{volume}{18}(\bibinfo{number}{1}), pp. \bibinfo{pages}{4--18},
  \doi{10.1145/321623.321625}.

\bibitemdeclare{book}{DBLP:books/sp/Greibach75}
\bibitem{DBLP:books/sp/Greibach75}
\bibinfo{author}{Sheila~A. \surnamestart Greibach\surnameend}
  (\bibinfo{year}{1975}): \emph{\bibinfo{title}{Theory of Program Structures:
  Schemes, Semantics, Verification}}.
\newblock {\sl \bibinfo{series}{Lecture Notes in Computer
  Science}}~\bibinfo{volume}{36}, \bibinfo{publisher}{Springer},
  \doi{10.1007/BFb0023017}.

\bibitemdeclare{article}{Immerman}
\bibitem{Immerman}
\bibinfo{author}{Neil \surnamestart Immerman\surnameend}
  (\bibinfo{year}{1988}): \emph{\bibinfo{title}{Nondeterministic Space is
  Closed Under Complementation}}.
\newblock {\sl \bibinfo{journal}{{SIAM} J. Comput.}}
  \bibinfo{volume}{17}(\bibinfo{number}{5}), pp. \bibinfo{pages}{935--938},
  \doi{10.1137/0217058}.

\bibitemdeclare{book}{Jones:97:ComputabilityComplexity}
\bibitem{Jones:97:ComputabilityComplexity}
\bibinfo{author}{Neil~D. \surnamestart Jones\surnameend}
  (\bibinfo{year}{1997}): \emph{\bibinfo{title}{Computability and complexity -
  from a programming perspective}}.
\newblock \bibinfo{series}{Foundations of computing series},
  \bibinfo{publisher}{{MIT} Press}, \doi{10.7551/mitpress/2003.001.0001}.

\bibitemdeclare{article}{logspace}
\bibitem{logspace}
\bibinfo{author}{Neil~D. \surnamestart Jones\surnameend}
  (\bibinfo{year}{1999}): \emph{\bibinfo{title}{{LOGSPACE} and {PTIME}
  Characterized by Programming Languages}}.
\newblock {\sl \bibinfo{journal}{Theor. Comput. Sci.}}
  \bibinfo{volume}{228}(\bibinfo{number}{1-2}), pp. \bibinfo{pages}{151--174},
  \doi{10.1016/S0304-3975(98)00357-0}.

\bibitemdeclare{article}{jfp}
\bibitem{jfp}
\bibinfo{author}{Neil~D. \surnamestart Jones\surnameend}
  (\bibinfo{year}{2001}): \emph{\bibinfo{title}{The expressive power of
  higher-order types or, life without {CONS}}}.
\newblock {\sl \bibinfo{journal}{J. Funct. Program.}}
  \bibinfo{volume}{11}(\bibinfo{number}{1}), pp. \bibinfo{pages}{55--94},
  \doi{10.1017/S0956796800003889}.
\newblock
  \urlprefix\url{http://journals.cambridge.org/action/displayAbstract?aid=68581}.

\bibitemdeclare{book}{Knuth:73:ArtOfProgramming}
\bibitem{Knuth:73:ArtOfProgramming}
\bibinfo{author}{Donald~E. \surnamestart Knuth\surnameend}
  (\bibinfo{year}{1973}): \emph{\bibinfo{title}{The Art of Computer
  Programming, Volume {III:} Sorting and Searching}}.
\newblock \bibinfo{publisher}{Addison-Wesley}.

\bibitemdeclare{inproceedings}{DBLP:conf/esop/KopS17}
\bibitem{DBLP:conf/esop/KopS17}
\bibinfo{author}{Cynthia \surnamestart Kop\surnameend} \&
  \bibinfo{author}{Jakob~Grue \surnamestart Simonsen\surnameend}
  (\bibinfo{year}{2017}): \emph{\bibinfo{title}{The Power of Non-determinism in
  Higher-Order Implicit Complexity - Characterising Complexity Classes Using
  Non-deterministic Cons-Free Programming}}.
\newblock In \bibinfo{editor}{Hongseok \surnamestart Yang\surnameend}, editor:
  {\sl \bibinfo{booktitle}{Programming Languages and Systems - 26th European
  Symposium on Programming, {ESOP} 2017, Held as Part of the European Joint
  Conferences on Theory and Practice of Software, {ETAPS} 2017, Uppsala,
  Sweden, April 22-29, 2017, Proceedings}}, {\sl \bibinfo{series}{Lecture Notes
  in Computer Science}} \bibinfo{volume}{10201}, \bibinfo{publisher}{Springer},
  pp. \bibinfo{pages}{668--695}, \doi{10.1007/978-3-662-54434-1\_25}.

\bibitemdeclare{article}{Kuroda}
\bibitem{Kuroda}
\bibinfo{author}{Sige-Yuki \surnamestart Kuroda\surnameend}
  (\bibinfo{year}{1964}): \emph{\bibinfo{title}{Classes of languages and
  linear-bounded automata}}.
\newblock {\sl \bibinfo{journal}{Information and Control}}
  \bibinfo{volume}{7}(\bibinfo{number}{2}), pp. \bibinfo{pages}{207--223},
  \doi{10.1016/S0019-9958(64)90120-2}.

\bibitemdeclare{article}{mccarthy}
\bibitem{mccarthy}
\bibinfo{author}{John \surnamestart McCarthy\surnameend}
  (\bibinfo{year}{1960}): \emph{\bibinfo{title}{Recursive Functions of Symbolic
  Expressions and Their Computation by Machine, Part {I}}}.
\newblock {\sl \bibinfo{journal}{Commun. {ACM}}}
  \bibinfo{volume}{3}(\bibinfo{number}{4}), pp. \bibinfo{pages}{184--195},
  \doi{10.1145/367177.367199}.

\bibitemdeclare{book}{moschovakis}
\bibitem{moschovakis}
\bibinfo{author}{Yiannis \surnamestart Moschovakis\surnameend}
  (\bibinfo{year}{2019}): \emph{\bibinfo{title}{Abstract Recursion and
  Intrinsic Complexity}}.
\newblock \bibinfo{publisher}{Cambridge University Press (Lecture Notes in
  Logic)}, \doi{10.1017/9781108234238}.

\bibitemdeclare{book}{papadimitriou}
\bibitem{papadimitriou}
\bibinfo{author}{Christos~H. \surnamestart Papadimitriou\surnameend}
  (\bibinfo{year}{1994}): \emph{\bibinfo{title}{Computational complexity}}.
\newblock \bibinfo{publisher}{Addison-Wesley}.

\bibitemdeclare{article}{DBLP:journals/jcss/Savitch70}
\bibitem{DBLP:journals/jcss/Savitch70}
\bibinfo{author}{Walter~J. \surnamestart Savitch\surnameend}
  (\bibinfo{year}{1970}): \emph{\bibinfo{title}{Relationships Between
  Nondeterministic and Deterministic Tape Complexities}}.
\newblock {\sl \bibinfo{journal}{J. Comput. Syst. Sci.}}
  \bibinfo{volume}{4}(\bibinfo{number}{2}), pp. \bibinfo{pages}{177--192},
  \doi{10.1016/S0022-0000(70)80006-X}.

\bibitemdeclare{article}{Szelepcsenyi}
\bibitem{Szelepcsenyi}
\bibinfo{author}{R{\'{o}}bert \surnamestart Szelepcs{\'{e}}nyi\surnameend}
  (\bibinfo{year}{1988}): \emph{\bibinfo{title}{The Method of Forced
  Enumeration for Nondeterministic Automata}}.
\newblock {\sl \bibinfo{journal}{Acta Inf.}}
  \bibinfo{volume}{26}(\bibinfo{number}{3}), pp. \bibinfo{pages}{279--284},
  \doi{10.1007/BF00299636}.

\bibitemdeclare{article}{turing}
\bibitem{turing}
\bibinfo{author}{Alan~M. \surnamestart Turing\surnameend}
  (\bibinfo{year}{1936-7}): \emph{\bibinfo{title}{On Computable Numbers with an
  Application to the {E}ntscheidungsproblem}}.
\newblock {\sl \bibinfo{journal}{Proceedings of the London Mathematical
  Society}} \bibinfo{volume}{42}(\bibinfo{number}{2}), pp.
  \bibinfo{pages}{230--265}, \doi{10.1112/plms/s2-42.1.230}.

\end{thebibliography}

\nocite{*}

\end{document}